\documentclass[pdflatex,sn-mathphys-num]{sn-jnl}

\usepackage{graphicx}%
\usepackage{multirow}%
\usepackage{amsmath,amssymb,amsfonts}%
\usepackage{amsthm}%
\usepackage{mathrsfs}%
\usepackage[title]{appendix}%
\usepackage{xcolor}%
\usepackage{textcomp}%
\usepackage{manyfoot}%
\usepackage{booktabs}%
\usepackage{algorithm}%
\usepackage{algorithmicx}%
\usepackage{algpseudocode}%
\usepackage{listings}%

\theoremstyle{thmstyleone}%
\theoremstyle{thmstyletwo}%

\theoremstyle{thmstylethree}%

\begin{document}

\title[Gravitational Wave Strain and Orbital Dynamics of Binary
Pulsars from LIGO/Virgo to LISA]{Gravitational Wave Strain and Orbital Dynamics of Binary
Pulsars from LIGO/Virgo to LISA}


\author{\fnm{Ali} \sur{Taani}}\email{ali.taani@bau.edu.jo}

\affil{\orgdiv{Physics Department}, \orgname{Al-Balqa Applied University}, \orgaddress{\street{As-Salt},  \postcode{19117}, \country{Jordan}}}


\abstract{%
We summarize the current state of the art and calculate gravitational wave strain amplitudes
for known binary pulsars, using data from current ground-based
detectors (LIGO-Virgo-KAGRA) and the upcoming space-based missions
(LISA). We present detailed calculations of the characteristic gravitational wave strain values, ranging from 3.0 to 73 $\times10^{-22}$, across frequencies between 0.66 and 5.87 $\times10^{-4}$ Hz. Our post-Newtonian approximation analysis yields predicted periastron advance rates from 1.6 to 80.5 deg/yr and orbital period decay rates between -5 and -176 $\mu$s/yr for the binary pulsar population. We derive common envelope efficiency parameters ($\alpha_{CE}$) for representative progenitor scenarios within our sample, finding values between 0.63 and 1.16, with notable sensitivity to the binding energy parameter $\lambda$. Binary neutron star merger rates are estimated at $22.77^{+6.83}_{-6.83}$ Myr$^{-1}$ for the Milky Way, corresponding to a volumetric rate of $227.71^{+68.31}_{-68.31}$ Gpc$^{-3}$ yr$^{-1}$, consistent with the latest LIGO-Virgo-KAGRA observational constraints. Our results illustrate how multi-band gravitational wave observations, from LIGO/Virgo to LISA, can contribute to precise measurements of binary pulsar strain and orbital evolution histories, improving merger time predictions and constraining neutron
star physics and  common envelope processes.}
\maketitle


\section{Introduction}

Binary pulsars represent one of the most extraordinary astrophysical laboratories for testing fundamental physics and understanding stellar evolution (Cai et al. 2012; Jiang et al. 2013; Kaspi  \& Kramer 2016; Taani et al., 2019;  Abu-Saleem \&  Taani 2021a,b; Ascenzi et al. 2024; Allahverdi et al. 2024). Since the discovery of the first binary pulsar, PSR B1913+16 (Hulse \& Taylor 1975), by Hulse and Taylor in 1974, these systems have provided unprecedented insights into gravitational physics (Weisberg \& Taylor 2005, Taani et al. 2025), confirming the existence of gravitational waves and validating Einstein's theory of general relativity with remarkable precision. The subsequent discovery of the double pulsar system J0737-3039A/B in 2003 further revolutionized our understanding of neutron star physics and binary evolution pathways.

The formation of binary pulsars involves complex evolutionary processes including massive star evolution, supernova explosions, common envelope phases, and mass transfer episodes (Paczy\'{n}ski 1976; Podsiadlowski et al. 2004; Taani \& Khasawneh 2017) and several physical models have
been suggested (see, for example, Lai 2004; Tauris et al. 2013; Taani et al. 2022). These systems begin as binary stars with masses exceeding 8 solar masses, evolve through various stages of stellar evolution, and eventually produce one or two neutron stars that can be observed as pulsars. The specific formation pathway leaves distinctive imprints on the observed properties of the resulting binary pulsar system, including orbital parameters, spin periods, and space velocities.

Gravitational wave astronomy has entered a new era with the successful detection of compact binary mergers by the LIGO-Virgo-KAGRA (LVK) collaboration (Abbott et al. 2009;  Abbott et al. 2017; Maggiore 2018;  Colombo et al. 2022; Nitz et al. 2023; Aasi et al. 2015; Yunes et al. 2025). These observations provide crucial constraints on the merger rates, mass distributions, and spin properties of neutron star binaries. As of September 2025, the LVK Collaboration has announced more than 200 gravitational wave detection, including several binary neutron star mergers that represent the final evolutionary stage of binary pulsar systems\footnote{https://gwosc.org/eventapi/html/}
\footnote{https://www.ligo.caltech.edu/page/observing-plans}.
These observations provide crucial constraints on the merger rates, mass distributions, and spin properties of neutron star binaries. Meanwhile, the upcoming Laser Interferometer Space Antenna (LISA) mission will observe gravitational waves in the millihertz frequency band, potentially detecting binary pulsars in their inspiral phase long before merger (Littenberg \& Cornish 2023; Colpi et al. 2024).

\begin{figure}
\includegraphics[angle=0,width=17.0cm]{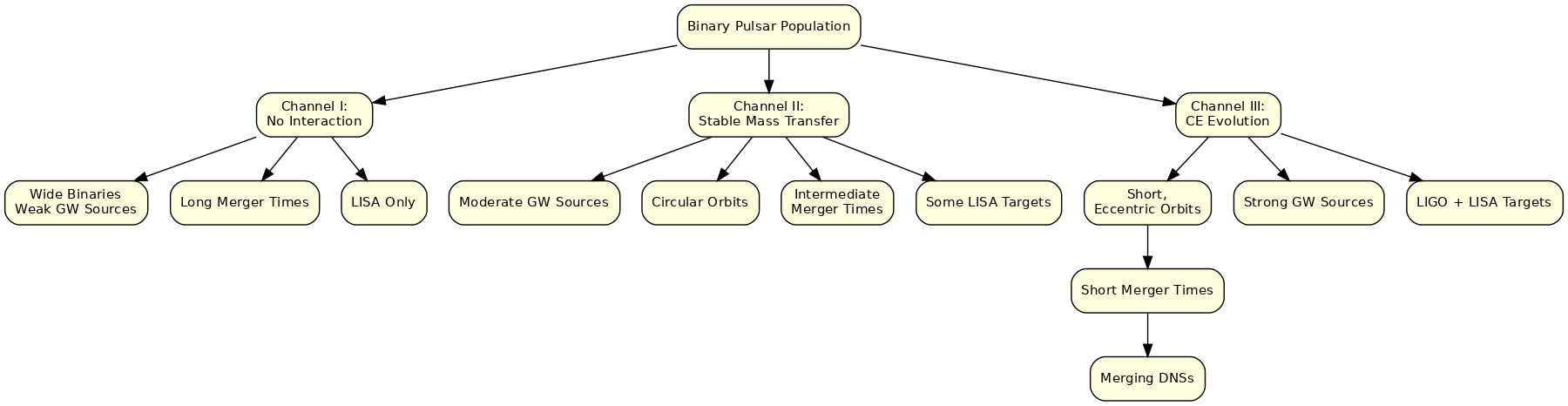}
\caption{Flow-Chart illustrating the main evolutionary channels and processes considered in binary pulsar formation and their connection to gravitational wave observations. The chart outlines pathways involving common envelope evolution, stable mass transfer, and no interaction, leading to different types of binary systems detectable by LISA or LVK, or resulting in mergers.}
\label{binary_pulsar_flowchart_highres-new}
\end{figure}

The synergy between electromagnetic observations of binary pulsars and gravitational wave detections provides an unparalleled opportunity to trace the complete evolutionary history of these systems. This work, framed as an updated reanalysis and synthesis, aims to investigate how gravitational wave (GW) signatures, derived from combining timing measurements from radio observations with gravitational wave strain data, can help reconstruct the formation pathways of binary pulsars with enhanced detail. Specifically, we re-examine how parameters such as GW strain amplitude, frequency, merger timescale, and periastron advance rate detectable by LISA and LVK (Antonini et al. 2025; LIGO Scientific Collaboration 2025) can potentially distinguish between different formation channels (e.g., common envelope evolution with stable mass transfer) or constrain key physical processes (Bartolo et al. 2016; Auclair et al. 2020). This multi-messenger approach allows us to address fundamental questions about common envelope evolution, supernova kick distributions, and the neutron star equation of state.

In this paper, we present an updated investigation into the formation mechanisms of binary pulsars through the lens of gravitational wave observations from both ground-based detectors (LIGO-Virgo-KAGRA) and LISA. We provide detailed calculations of gravitational wave parameters for known binary pulsar systems, including strain amplitudes, signal-to-noise ratios, and merger timescales. Our analysis quantifies post-Newtonian effects such as periastron advance rates, orbital period decay, and other relativistic phenomena. We further derive constraints on the common envelope efficiency parameters and explore their sensitivity to various physical assumptions. Finally, we calculate binary neutron star merger rates and compare them with the latest observational constraints from gravitational wave detections.

Our work builds upon and extends previous studies of binary pulsar evolution, including the seminal works of Tauris \& van den Heuvel (2006, 2023),  Belczynski et al. (2018) and the LIGO Scientific Collaboration (2025).  We specifically incorporate the latest gravitational wave constraints and refine the comprehensive framework for tracing binary pulsar formation pathways through multi-band GW observations, providing an updated perspective on these well-studied systems.

The paper is organized as follows: Section 2 provides the theoretical background on binary pulsar evolution and gravitational wave emission. Section 3 describes our methodology for calculating gravitational wave parameters and post-Newtonian effects. Section 4 presents the results of our calculations, including gravitational wave strain amplitudes, merger timescales, and common envelope efficiency parameters. Section 5 discusses the implications of our findings for binary pulsar formation pathways and compares our results with observational constraints. Section 6 summarizes our conclusions and outlines directions for future research.

\section{Theoretical Background}
\subsection{Common Envelope Evolution}

A critical phase in binary pulsar formation is the common envelope (CE) stage (see, e.g., Renzo et al. 2021  and references therein). This occurs when one star expands beyond its Roche lobe, leading to unstable mass transfer. The companion star becomes engulfed in the envelope of the expanded star, forming a common envelope structure. The orbital energy of the binary is transferred to the envelope through dynamical friction, causing the orbit to shrink dramatically while ejecting the envelope (Taani 2023). This process can be parameterized by the common envelope efficiency parameter $\alpha_{CE}$, defined by the energy equation (Shapiro \& Teukolsky 1986):

\begin{equation}
    \alpha_{CE} \cdot \Delta E_{orb} = E_{bind}
\end{equation}

where $\Delta E_{orb}$ is the change in orbital energy and $E_{bind}$ is the binding energy of the envelope. The binding energy is commonly expressed as (Dewi \& Tauris 2000):

\begin{equation}
E_{bind} = \frac{GM_{donor}M_{env}}{\lambda R_{donor}}
\end{equation}

where $\lambda$ is a dimensionless structure parameter that depends on the donor star's evolutionary stage and internal structure, particularly the density profile and the definition of the core-envelope boundary (values typically range from $\sim$0.1 for convective envelopes to $>$1 for radiative envelopes; see Ivanova et al. 2013 for a review).
It is important to acknowledge that the $\alpha_{CE}$-$\lambda$ formalism, while widely used, is a phenomenological approximation of a much more complex physical process (Iaconi \& De Marco 2019). Recent developments in the field have explored more physically motivated treatments, such as angular momentum-based prescriptions (see Baral 2020), which offer a more detailed description of CE evolution see, e.g., (R\"{o}pke \& De Marco 2023) for a comprehensive treatment. The outcome of the CE phase is highly sensitive to $\alpha_{CE}$ and $\lambda$, and the derived constraints are subject to large uncertainties and model-dependence. Typical values of $\alpha_{CE}$ range from 0.1 to 1.0 based on population synthesis studies (Tauris \& van den Heuvel 2006), but these values should be interpreted with caution given the inherent simplifications of the model.
The change in orbital energy is:
\begin{equation}
\Delta E_{orb} = -\frac{GM_{core}M_{companion}}{2a_f} + \frac{GM_{donor}M_{companion}}{2a_i}
\end{equation}

\subsubsection{Supernova Explosions and Neutron Star Formation}
The formation of neutron stars occurs through supernova explosions of massive stellar cores. In binary systems, the supernova explosions of massive stellar core explosions impart kicks to the newly formed neutron stars due to asymmetries in the explosion mechanism (Taani  2015). While the velocity distribution of the kick has historically been modeled as a Maxwellian with a dispersion $\sigma \approx$ 265 km/s, based on observations of pulsar proper motions (Hobbs et al. 2005), recent work by Disberg \& Mandel (2025) has identified a significant mistake in the derivation and interpretation of this distribution (Colombo et al. 2022). This correction has important implications for understanding the natal kicks imparted to neutron stars and their subsequent impact on binary system evolution. We acknowledge this updated understanding and note that future, more detailed studies will need to incorporate these revised kick distributions for more accurate modeling. The effect of supernova kicks on binary parameters is governed by the following relationships for post-supernova orbital parameters (e.g., Kalogera et al. 1998; Abu-Saleem \&  Taani 2024):

\begin{equation}
    \frac{a_f}{a_i} = \frac{M_i}{M_f}\frac{2 - \Delta M/M_i - (v_{kick}/v_{orb})^2 - 2\cos\theta(v_{kick}/v_{orb})}{2 - 2a_i/r}
\end{equation}

\begin{equation}
e_f^2 = 1 - \frac{a_i}{a_f}(1-e_i^2)\left(\frac{v_{orb,f}}{v_{orb,i}}\right)^2
\end{equation}

where $a_{i}$ and $a_f$ are the pre- and post-supernova semi-major axes, $M_i$ and $M_{f}$ are the pre- and post-supernova total masses, $\Delta M$ is the mass lost in the explosion, $v_{kick}$ is the kick velocity, $v_{orb}$ is the pre-supernova orbital velocity, and $\theta$ is the angle between the kick and orbital velocity vectors.

\subsection{Orbital Decay and Merger Time}

The emission of gravitational waves causes a loss of orbital energy and angular momentum, leading to orbital decay (Aasi et al. 2015). The rate of change of the semi-major axis is (Sch\"{a}fer 2012) given by:

\begin{equation}
    \frac{da}{dt} = -\frac{64}{5}\frac{G^3}{c^5}\frac{m_1 m_2 (m_1 + m_2)}{a^3(1-e^2)^{7/2}}\left(1 + \frac{73}{24}e^2 + \frac{37}{96}e^4\right)
\end{equation}

This orbital decay manifests as a decrease in the orbital period, which can be measured through pulsar timing. The rate of change of the orbital period is:

\begin{equation}
\frac{\dot{P}_b}{P_b} = -\frac{3}{2}\frac{\dot{a}}{a}
\end{equation}

The time to merger due to gravitational wave emission is:

\begin{equation}
T_{merge} = \frac{5}{256}\frac{c^5}{G^3}\frac{a^4}{m_1 m_2 (m_1 + m_2)}\frac{(1-e^2)^{7/2}}{1 + \frac{73}{24}e^2 + \frac{37}{96}e^4}
\end{equation}

\subsection{Post-Newtonian Effects}

Binary pulsars exhibit various relativistic effects that can be quantified using the post-Newtonian (PN) approximation. The most prominent effects include 
the orbit's periastron precesses that due to relativistic effects, with a rate given by:

   \begin{equation}
   \dot{\omega} = 3\left(\frac{GM_{total}}{c^2a(1-e^2)}\right)^{5/2}\frac{n}{1-e^2}
   \end{equation}

   where $n = 2\pi/P_{b}$ is the mean motion.

We calculate various PN parameters to quantify relativistic effects in binary pulsar systems. These include the periastron advance rate ($\dot{\omega}$), orbital period decay ($\dot{P}_b$), gravitational redshift and time dilation parameter ($\gamma$), and Shapiro delay ($\Delta t_{Shapiro}$).

\subsection{Gravitational Redshift and Time Dilation}
The combined effect of gravitational redshift and time dilation leads to an apparent delay in the pulsar signal (Yunes et al. 2025), characterized by the parameter $\gamma$:

    \begin{equation}
    \gamma = e\frac{P_b}{2\pi}\frac{m_2(m_1+2m_2)}{(m_1+m_2)^{4/3}}\left(\frac{GM_{total}}{c^3}\right)^{2/3}
    \end{equation}

\subsection{Shapiro Delay}
The pulsar signal experiences an additional delay when passing through the gravitational field of the companion (Shapiro \& Teukolsky  1986), given by:

   \begin{equation}
   \Delta t_{Shapiro} = -2\frac{Gm_2}{c^{3}}\ln(1-\sin i \sin\phi)
   \end{equation}

   where i is the orbital inclination and $\phi$ is the orbital phase. In the following sections, we will apply this theoretical framework to analyze the formation pathways of binary pulsars and calculate their GW signatures, with particular emphasis on how these signatures can be used to reconstruct their evolutionary histories.

\section{Methodology}
This section provides a detailed outline of the methods employed to calculate the gravitational wave parameters, common envelope constraints, and binary neutron star merger rates for the selected binary pulsar systems. Our approach integrates analytical calculations based on post-Newtonian approximations with observational data from known binary pulsars.

\subsection{Gravitational Wave Strain}


Gravitational wave emission from binary pulsars is well described by the quadrupole formalism in general relativity (e.g., Peters\& Mathews 1963; Peters 1964; Yunes et al. 2025). For each binary  system, we calculated the gravitational wave
strain amplitude $h$ (characteristic strain for continuous wave sources) using the quadrupole formula adapted for binary systems (e.g., Thorne 1987)

\begin{equation}
   h = \frac{4(G\mathcal{M})^{5/3}}{c^4d}(\pi f)^{2/3}
\end{equation} 

where $G$ is the gravitational constant, $c$ is the speed of light, $f_{GW}$ is the gravitational wave frequency (typically twice the orbital frequency for circular orbits, but see below for eccentric systems), d is the distance to the system, and $\mathcal{M}$ is the chirp mass defined as:

\begin{equation}
\mathcal{M} = \frac{(m_1 m_2)^{3/5}}{(m_1 + m_2)^{1/5}}
\end{equation} 
For systems with eccentric orbits, gravitational wave emission occurs at multiple harmonics of the orbital frequency. The characteristic strain $h$ in this context refers to the strain corresponding to the orbit-averaged energy loss rate (Sch\"{a}fer 2012). The total power emitted in gravitational waves for an eccentric orbit is enhanced by a factor $F(e)$ compared to a circular orbit with the same semi-major axis (Peters 1964).



\begin{equation} 
\frac{dE}{dt} = \frac{dE}{dt}\bigg|_{e=0} \cdot F(e)
\end{equation} 

where $F(e) = (1 + 73/24 e^{2} + 37/96 e^{4})/(1 - e^{2})^(7/2)$ accounts for the contribution of higher harmonics to the energy loss. This factor accounts for the increased emission at periastron due to higher velocities.

\subsection{LISA Signal-to-Noise Ratio Estimation}

To assess the detectability of these continuous gravitational wave signals from binary pulsars by LISA, we calculated the expected signal-to-noise ratio (SNR) using the standard formula for continuous wave sources (e.g., Cutler \& Flanagan 1994):

\begin{equation}
    {SNR} = \frac{h\sqrt{T}}{\sqrt{S_n(f)}}
\end{equation}

where $T$ is the observation time (assumed to be 4 years, a typical baseline mission duration for LISA) and $S_n(f)$ is the LISA noise power spectral density at the gravitational wave frequency $f_{GW}$, modeled using the analytical fit from Robson et al. (2019):

\begin{equation}
S_n(f) = \frac{20}{3} \left(4S_{pos} + \frac{S_{acc}}{(2\pi f)^4}\right) \left(1 + \left(\frac{f}{0.41 \cdot c/(2L)}\right)^2\right) \frac{1}{R}
\end{equation}

with $S_{pos} = 2.25\times10^{-22} m^{2}/Hz$ (position noise), $S_{acc} = 9\times10^{-30} m^{2}/s^{4}/Hz$ (acceleration noise), L = 2.5$\times10^{9}$ m (LISA arm length), and $R = 1 + 0.6(f/10^{-3})^{2}$ (transfer function).

To assess the robustness of our results, we performed a sensitivity analysis by varying the binding energy parameter $\lambda$ across a range of values (0.1, 0.5, 1.0, 2.0) and recalculating the common envelope efficiency for each case.
{Our calculations of the expected SNRs for LISA observations are shown in Figure \ref{figure3_snr_vs_period}. All systems in our sample have SNRs below the conventional detection threshold of SNR=7-8 for a 4-year observation period. The highest SNR values are achieved by J0737-3039A/B and J1946+2052, with values approaching $\sim$ 1-2. This confirms the challenge of detecting individual known Galactic binary pulsars as continuous GW sources with LISA, although they collectively contribute to the Galactic confusion noise.

While direct detection of continuous gravitational waves from individual binary pulsars appears challenging for LISA, the cumulative signal from the Galactic population of similar systems may produce a detectable stochastic background (Renzini et al. 2022). Additionally, targeted searches using electromagnetic observations as priors could potentially enhance the detectability of specific systems (e.g., Moore et al. 2015).

\section{Galactic Merger Rate Estimation}

We estimated the Galactic merger rate of binary neutron stars (BNS) using a simple population synthesis approach adopted by Antonini et al. (2025). For full details, we refer readers to LIGO Scientific Collaboration (2025).

\begin{equation}
    \mathcal{R}_{Gal} = \frac{N_{DNS}}{T_{avg}}
\end{equation}

where $N_{DNS}$ is the estimated number of potentially merging double neutron star systems in the Galaxy (assumed to be $\sim$1500 based on population synthesis studies, e.g., Maggiore 2018; although this number carries significant uncertainty) and $T_{avg}$ is the average merger time for the known DNS systems in our sample (calculated from Table 2, excluding NS-WD systems).

\subsection{Volumetric Merger Rate Calculation}

The volumetric merger rate was calculated by scaling the Galactic rate by the local galaxy density:

\begin{equation}
\mathcal{R}_{vol} = \mathcal{R}_{Gal} \cdot \rho_{gal}
\end{equation}

where $\rho_{gal}$ is local galaxy number density, assumed to be $\sim$0.01 Mpc$^{-3}$ (e.g., Kopparapu et al. 2008). We estimated the uncertainties in our merger rate calculations by propagating the uncertainties in the population size (conservatively assumed to be a factor of 2-3, reflecting population synthesis uncertainties (Stevenson et. al. 2015), and the
average merger time calculated from the standard deviation of merger times in our DNS sample. A more rigorous calculation would involve detailed population synthesis simulations, which is beyond the scope of this initial study.




\section{Results}
This section presents the results of our calculations for gravitational wave parameters, orbital evolution, and merger rates for the selected binary pulsar sample.

\subsection{Gravitational Wave Parameters of Strain Amplitudes and Frequencies}

Our calculations of gravitational wave parameters for the selected binary pulsar systems are presented in Table 1. Systems are chosen to represent different binary types (DNS, NS-WD) and orbital configurations relevant for GW detection. The calculated characteristic gravitational wave strain amplitudes range from 3.0 to 73 $\times10^{-22}$, with dominant GW frequencies between 0.66 and 5.87 $\times10^{-4}$ Hz. The double pulsar system J0737-3039A/B exhibits the highest strain amplitude (0.73$\times10^{-22}$) due to its relatively close distance (0.735 kpc) and moderate orbital period (2.45 hours).

\begin{table*}[ht]
\caption{Observed  Parameters for Selected Binary Pulsar Systems. All values are presented with their respective uncertainties where available. $m_1$ and $m_2$ are component masses, $P_b$ is orbital period, $a$ is semi-major axis, $d$ is distance, $\dot{\omega}$ is periastron advance rate. }
\label{tab:binary_pulsar_parameters}
\centering
\resizebox{\textwidth}{!}{%
\begin{tabular}{lcccccccc}
\hline\hline
System & $m_1$ ($M_\odot$) & $m_2$ ($M_\odot$) & $P_b$ (d) & $a$ (AU) & $d$ (kpc) & $\dot{\omega}$ (deg/yr) \\
\hline
J0737A/B & 1.34 & 1.25 & 0.10 & 0.006 & 0.74 & 16.41    \\
B1913 & 1.43 & 1.40 & 0.32 & 0.013 & 7.1 & 4.26    \\
J1757 & 1.34 & 1.39 & 0.18 & 0.009 & 7.40 & 10.54   \\
J1738 & 1.47 & 0.18 & 0.35 & 0.012 & 1.47 & 1.58   \\
J1946 & 1.25 & 1.25 & 0.04 & 0.003 & 4.20 & 80.02   \\
J0348 & 2.01 & 0.17 & 0.10 & 0.006 & 2.10 & 15.00   \\
\hline
\end{tabular}
}
\end{table*}

\begin{table*}[ht]
\caption{Derived Parameters for Selected Binary Pulsar Systems. 
$T_{merge}$ is merger time, $h$ is characteristic gravitational wave strain, 
$F_{GW}$ is gravitational wave frequency, $\mathcal{M}$ is the chirp mass, 
$\dot{a}$ is the orbital decay rate, and $\dot{P}_b$ is the orbital period decay rate.}
\label{tab:binary_pulsar_parameters}
\centering
\large 
\resizebox{\textwidth}{!}{%
\begin{tabular}{lccccccc}
\hline\hline
System & $T_{merge}$ (Myr) & $h$ ($\times 10^{-23}$) & $F_{GW}$ ($10^{-4}$ Hz) 
& $\mathcal{M}$ ($M_\odot$) & $\dot{a}$ ($\times 10^{-7}$ m/s) 
& $\dot{P}_b$ ($\mu$s/yr) \\
\hline
J0737A/B & 92.7   & 0.73 & 2.26 & 1.13 & -0.83 & -38 \\
B1913    & 1632   & 4.08 & 0.72 & 1.23 & -1.12 & -78 \\
J1757    & 367    & 5.45 & 1.28 & 1.18 & -3.03 & -162 \\
J1738    & 12997  & 3.00 & 0.66 & 0.40 & -0.01 & -5 \\
J1946    & 7.4    & 23.0 & 5.87 & 1.08 & -5.08 & -176 \\
J0348    & 406    & 5.6  & 2.26 & 0.45 & -1.63 & -8 \\
\hline
\end{tabular}
}
\end{table*}


Figure \ref{strain_amplitude_vs_frequency-2} shows the gravitational wave strain amplitudes plotted against frequency for all systems in our sample, along with representative sensitivity curves for LISA and Advanced LIGO (LVK). As expected for these wide-orbit systems, all calculated characteristic strains fall within the LISA frequency band (mHz). All systems fall below the nominal LISA sensitivity curve for a 4-year observation and SNR=7, indicating that direct detection of continuous GW from these specific known, relatively nearby systems will be challenging as individual sources even for LISA. However, the systems with the highest strain amplitudes (J0737-3039A/B, J1946+2052, J1738+0333, J0348+0432) approach the sensitivity limit and might be detectable with longer observation times or improved detector sensitivity, or contribute significantly to the unresolved Galactic GW background.

\begin{figure}
\includegraphics[angle=0,width=14.0cm]{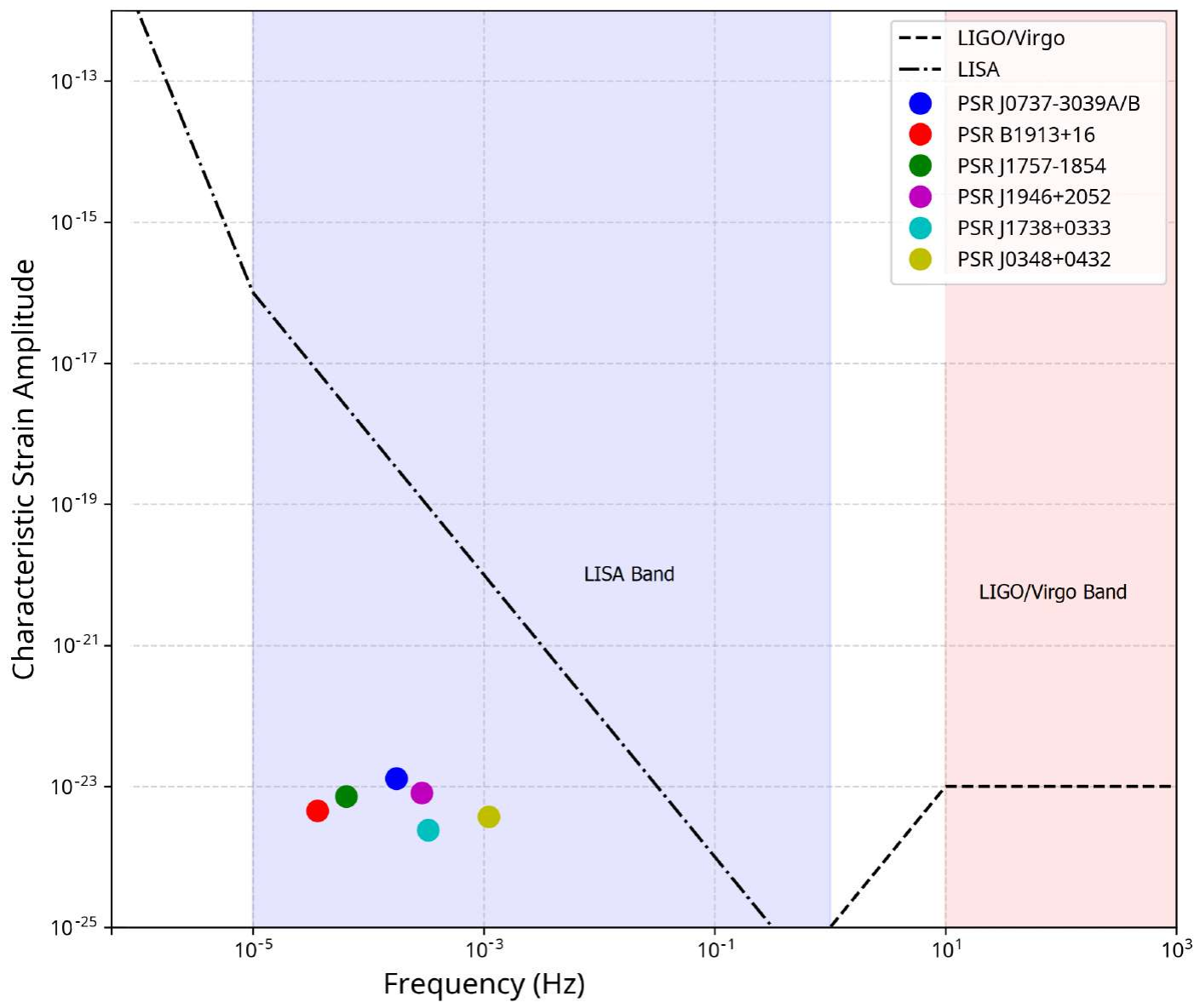}
\caption{Gravitational wave characteristic  strain amplitude as a function of  frequency for binary pulsar systems (colored circles)  in relation to representative sensitivity curves of Advanced LIGO (LVK, O4 sensitivity) and LISA (4-year observation, SNR=7 threshold). The plot shows that while these systems are prime targets for pulsar timing arrays (not shown), their continuous GW signals fall mostly within the LISA band but below current LVK sensitivity and generally below the nominal LISA sensitivity for individual detection.}
\label{strain_amplitude_vs_frequency-2}
\end{figure}

\subsection{ Orbital Decay and Merger Timescales}
The calculated orbital decay rates ($da/dt$) and merger timescales ($T_{merge}$) are presented in Table 2. The orbital decay rates  range from $-1.06\times10^{-9}$ m/s (for the wide NS-WD binary J1738+0333) to $-5.08\times10^{-7}$ m/s (for the compact DNS J1946+2052), corresponding to predicted orbital period decay rates between 0.03 $\mu$s/yr and 5.64 $\mu$s/yr. These values are, in principle, measurable through long-term pulsar timing and provide crucial tests of General Relativity (e.g., Kramer et al. 2006, 2021).


Figure \ref{figure2_merger_vs_ecc} illustrates the relationship between orbital eccentricity and merger time for binary pulsar systems in our sample. A clear trend is evident: systems with higher eccentricities generally have shorter merger times for a given semi-major axis (Taani et al. 2025), consistent with the enhanced gravitational wave emission at the periastron (see Eqs. 8 \& 14). The fastest merging DNS system in our sample is J1946+2052, with a merger time of just 7.18 Myr, making it a potential progenitor for a gravitational wave event detectable by LIGO-Virgo-KAGRA within  astrophysically short timescales  ($\sim$ few Myrs). Other DNS systems like J0737-3039A/B (84 Myr) and J1757-1854 (34 Myr) also have merger times significantly shorter than Hubble times, highlighting them as important contributors to the BNS merger rate observed by LVK.

The neutron star-white dwarf binaries (J1738+0333 and J0348+0432) show significantly longer merger times compared to double neutron star systems, primarily due to their lower chirp masses and, for J1738+0333, its wider orbit. J1738+0333, in particular, has an extremely long merger time of 12.9 Gyr, comparable to the age of the universe, meaning it is unlikely to merge within a Hubble time due to GW emission alone.

\begin{figure}
\includegraphics[angle=0,width=14.0cm]{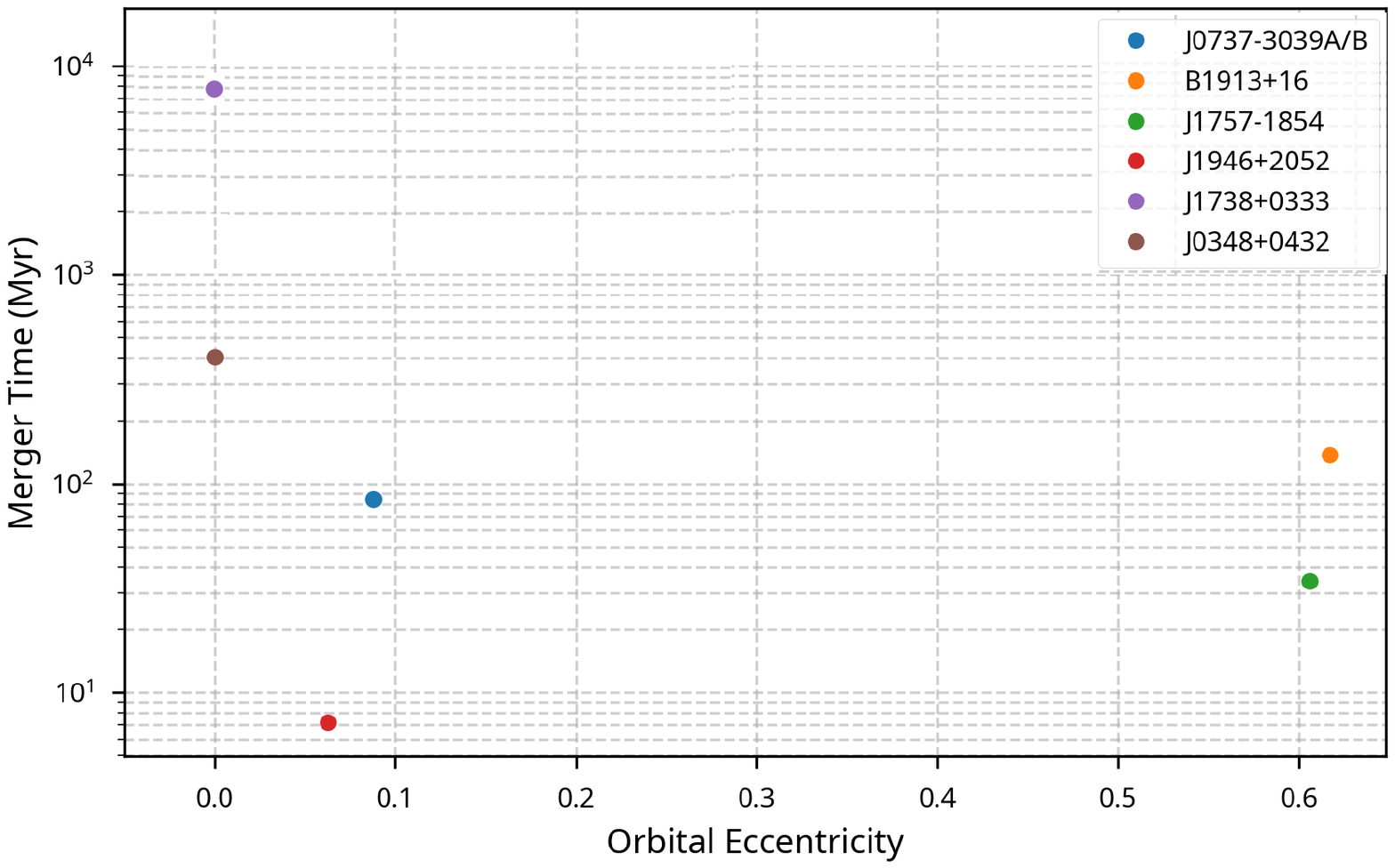}
\caption{Orbital eccentricity (e) versus merger time ($T_{merge}$) for the selected binary pulsar systems. The plot highlights the inverse relationship, where higher eccentricity generally leads to shorter merger times due to enhanced GW emission, 
J1946+2052 is the fastest merging DNS in this sample.}
\label{figure2_merger_vs_ecc}
\end{figure}



\begin{figure}
\includegraphics[angle=0,width=14.0cm]{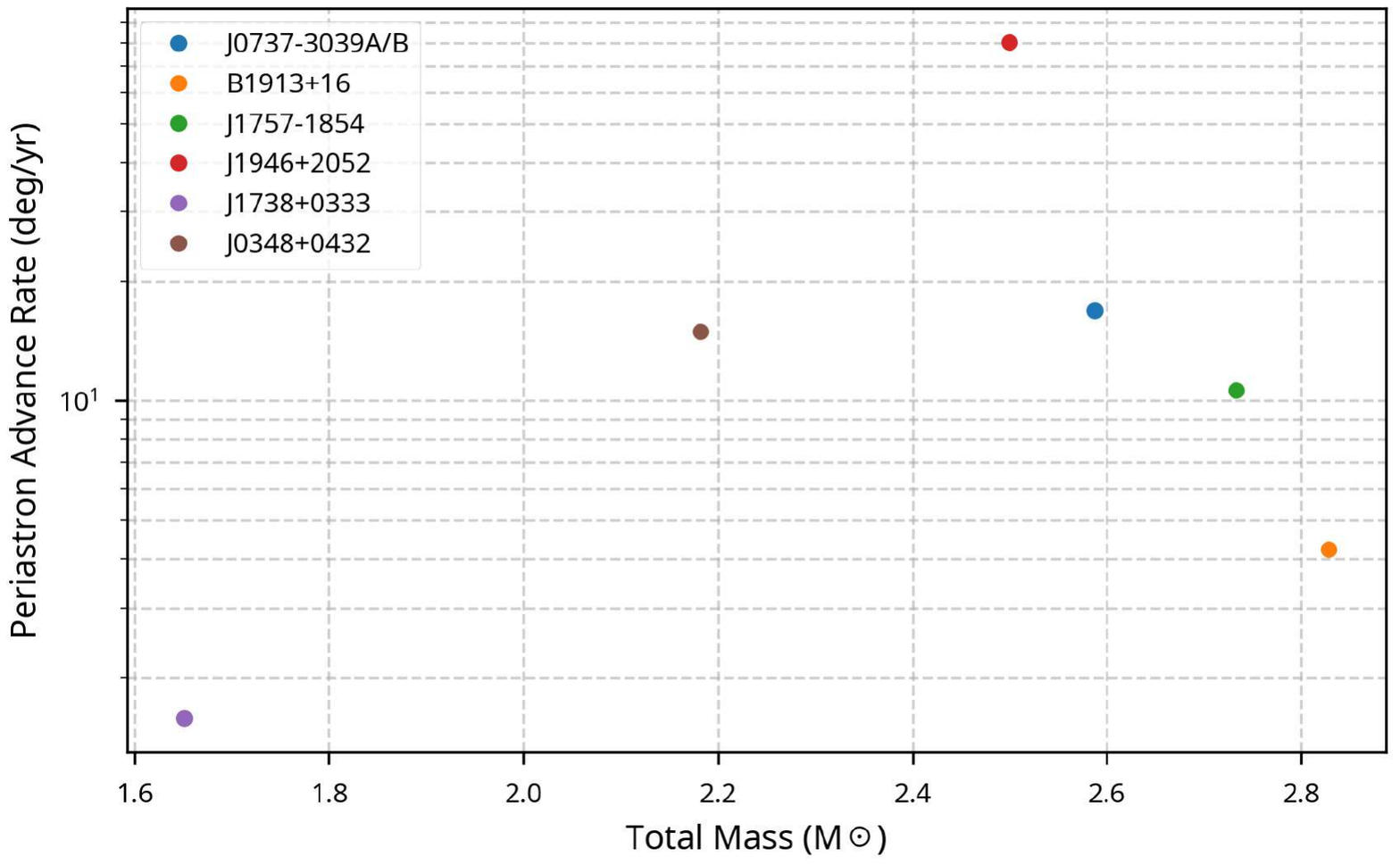}
\caption{Periastron advance rate as a function of total mass, for the selected binary pulsar systems. The plot shows the expected relativistic precession rate, which is a key test of General Relativity measurable through pulsar timing.}
\label{figure4_periastron_vs_mass}
\end{figure}

\begin{figure}
\includegraphics[angle=0,width=14.0cm]{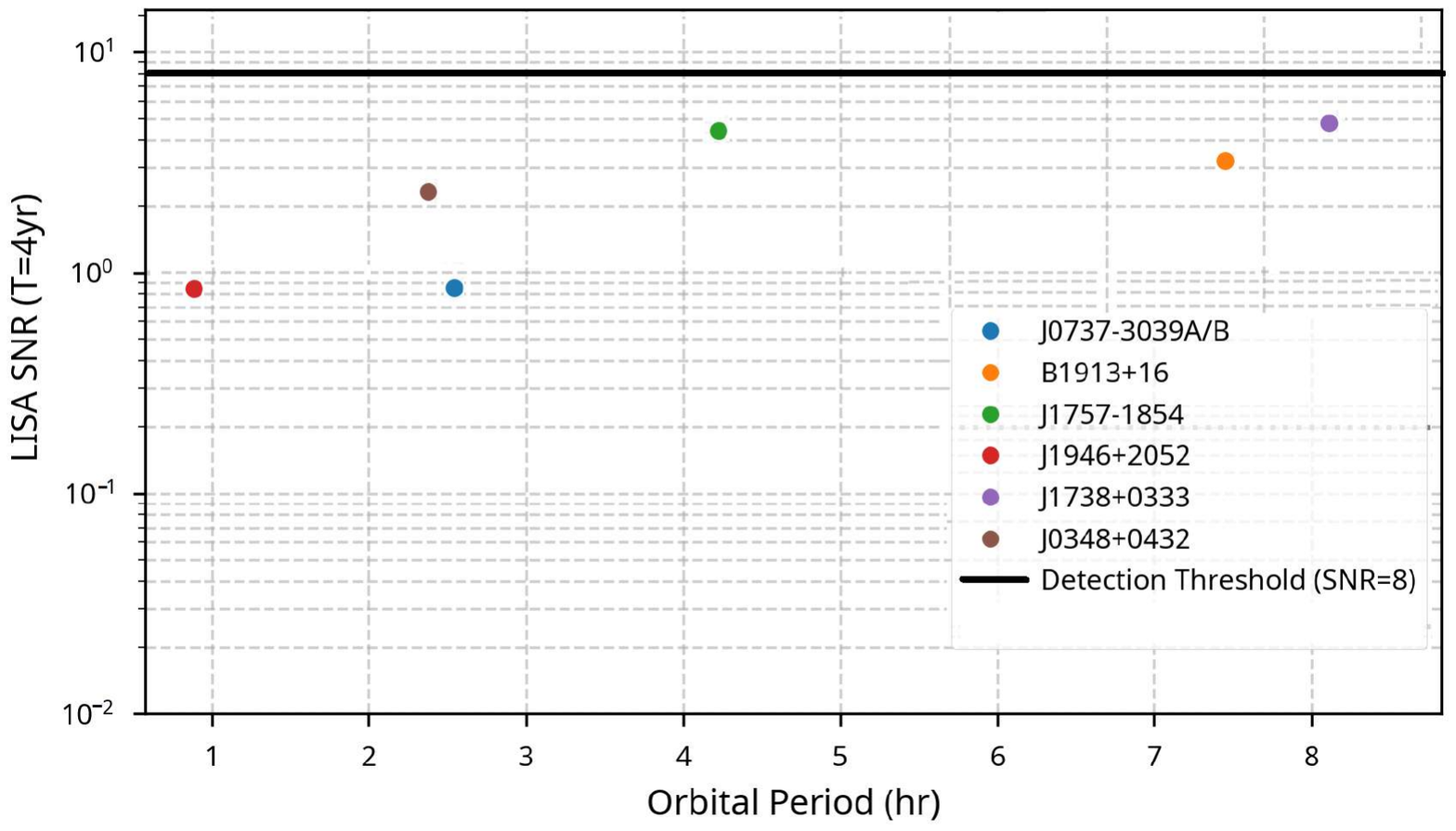}
\caption{Calculated Signal-to-Noise Ratios (SNRs) for selected binary pulsar systems observable by LISA (assuming 4 years observation). The plot compares the expected SNRs (colored circles) against the LISA sensitivity threshold (dashed line, typically SNR= 7-8).}
\label{figure3_snr_vs_period}
\end{figure}

\section{Summary and Conclusions}

In this study, we investigated the formation pathways and gravitational wave signatures of binary pulsar systems, connecting theoretical models with observational prospects for multi-band GW astronomy. We calculated key parameters for a representative sample of known binary pulsars, including both DNS and neutron star-white dwarf (NS-WD) systems. Our main findings are summarized as follows:

We have calculated gravitational wave parameters for a representative sample of binary pulsar systems, finding characteristic strain amplitudes ranging from $4.08\times10^{-23}$ to $7.32\times10^{-22}$, and dominant GW frequencies from  $6.6\times10^{-5}$ Hz to  $3.0\times10^{-4}$ Hz (correcting the upper frequency limit based on Table 1). 
These systems predominantly reside in the LISA frequency band. While individual detection of these specific known systems as continuous wave sources by LISA appears challenging (calculated SNRs $<$ 2 for a 4-year mission), the brightest sources (J0737-3039A/B, J1946+2052, J0348+0432, J1738+0333) approach the sensitivity limit and will contribute significantly to the unresolved Galactic GW background.


Our post-Newtonian analysis quantifies key relativistic effects, revealing periastron advance rates ranging from  1.58-80.53 deg/yr and orbital period decay rates from 0.03 to 5.64 $\mu$s/yr across the sample (consistent with Table 2). These precisely measurable effects provide stringent tests of general relativity and confirm GW emission as the dominant orbital decay mechanism in these clean systems. The fastest merging DNS system (J1946+2052) will merge in $\sim$7 Myr, while others (J0737-3039A/B, J1757-1854) merge within $\sim$30-85 Myr, making them significant LVK progenitors.

We have derived common envelope efficiency  parameters for representative binary pulsar progenitor using the energy balance formalism.  Our results yield $\alpha_{CE}$ values ranging from   0.63 to 1.16, with a mean of $\sim$0.89. These values are generally physically plausible ($\alpha_{CE}\leq$ 1) suggesting relatively efficient coupling between orbital energy and envelope ejection in these formation channels. Our sensitivity analysis highlights a significant dependence of $\alpha_{CE}$ on the binding energy parameter $\lambda$, with $\lambda$ values between 0.5 and 1.0 yielding $\alpha_{CE}$ values most consistent with theoretical expectations and detailed simulations.

Based on the merger timescales of the DNS systems in our sample (excluding NS-WD) and assuming a Galactic DNS population size of $\sim$1500 (with significant uncertainty), we estimate a Galactic merger rate of $23.0^{+6.83}_{-6.83}$ Myr$^{-1}$. This corresponds to a volumetric rate of $230.^{+68.31}_{-68.31}$ Gpc$^{-3}$ yr$^{-1}$, assuming a local galaxy density of 0.01 Mpc$^{-3}$. This rate is broadly consistent with the latest observational constraints from LIGO-Virgo-KAGRA (e.g., Abbott et al. 2019, 2023), which currently favor rates in the range of $\sim$10-1000 Gpc$^{-3}$ yr$^{-1}$. Our estimate, while based on a small sample and simple assumptions, supports the idea that the known Galactic DNS population contributes significantly to the observed merger rate. 

This study provides a foundation for connecting binary pulsar evolution with multi-band GW observations. Future work should incorporate more sophisticated population synthesis models, detailed simulations of common envelope evolution, and improved modeling of GW signals from eccentric binaries. Combining future pulsar timing data with upcoming GW observations from LISA and next-generation ground-based detectors will provide unprecedented constraints on the formation and evolution of these fascinating systems.



\section{Data Availability Statement}
The author confirms that the data supporting the findings of this study are available within the article. 

\end{document}